\documentclass{article}%
\usepackage{amsmath}
\usepackage{amsfonts}
\usepackage{amssymb}
\usepackage{graphicx}%
\begin{document}

\title{Energy Distribution of a Regular Class of Exact Black Hole Solutions}
\author{I-Ching Yang$^{\text{1}}$, Chi-Long Lin$^{\text{2}}$, I. Radinschi$^{\text{3}%
}$\\$^{\text{1}}$Department of Applied Science\\and Systematic and Theoretical Science Research Group,\\National Taitung University, Taitung, Taiwan 950,\\icyang@nttu.edu.tw,\\$^{\text{2}}$The National Museum of Natural Science,\\Taichung, Taiwan 403, Republic of China,\\and $^{\text{3}}$Department of Physics, "Gh. Asachi" Technical University,\\Iasi, 700050, Romania\\radinschi@yahoo.com}
\maketitle

\begin{abstract}
In this paper we present the expressions for the energy of a regular class
of exact black hole solutions of Einstein's equations coupled with a nonlinear
electrodynamics source. We calculate the energy distribution using the
Einstein, Weinberg and M\o ller prescriptions. We make a discussion of the
results in function of two specific parameters, a sort of dipole and
quadrupole moments of the nonlinear source $\alpha$ and $\beta$, and in
addition a study of some particular cases is performed.

\end{abstract}

\section{INTRODUCTION}

Energy-momentum localization has presented a remarkable progress in the recent
years, although a satisfactory generally accepted expression for energy
distribution has not been found. As an illustration, we notice the
pseudotensorial definitions [1]-[7] which have been used for computing the
energy of ~$3+1$, $2+1$ and $2$ dimensional geometries enlightening that
different energy-momentum complexes can yield the same expression for energy
distribution of a given space-time [8]-[9]. These nontensorial quantities
contain the contribution of the matter and non-gravitational fields and that
of the gravitational field, and give rise of the problem of coordinate
dependence. Among these energy-momentum prescriptions only the M\o ller's
definition can be applied to any coordinate system [10]. Although the
coordinate dependence, recently many interesting studies have been performed
for exploring various geometries and making more accessible the issue of
energy-momentum localization [8]-[10]. This involves understanding the
pseudotensorial mechanism, how can be used for the evaluation of energy of
different space-times and attempt to provide many illustrative examples. For
the Einstein, Landau-Lifshitz, Papapetrou and Weinberg prescriptions the
problem of coordinate dependence was partially solved in the case of the
metrics of Kerr-Schild class [11]. Further, for a great number of space-times
the energy-momentum complexes give the same results with their tele-parallel
versions [12]. Chang, Nester and Chen in an interesting work [13] concluded
that energy-momentum complexes are directly connected to quasilocal
expressions for the energy-momentum, and each energy-momentum complex is
associated with a legitimate Hamiltonian boundary term. The connection with
the boundary conditions also implies that each expression for energy has a
geometrically and physically significance. All these considerations point out
the significance of the energy-momentum complexes and stress their usefulness
for the localization of energy. Even the researchers are confident that the
pseudotensorial mechanism gives meaningful result, the main question remains
available, how the coordinate dependence could be overcome and a generally
consistent formula for energy-momentum density\ developed. Moreover, for
clarifying some weakness of the pseudotensorial definitions and point out
their best properties future work is needed.

In this paper we evaluate the energy of a regular class of exact black hole
solutions of Einstein's equations coupled with a nonlinear electrodynamics
source [14] using the Einstein, Weinberg and M\o ller prescriptions. We show
that the expression for the Weinberg covariant energy is equal to the
expression for the Einstein energy. The discussion will involve some
particular cases obtained in function of two specific parameters, and in
addition presents a study of some particular solutions connected with a
previous work [15].

The paper is organized as \ follows: in Section 2 we present the regular class
of four parametric exact black hole solutions obtained by Ay\'{o}n-Beato and
Garcia, the Einstein, Weinberg and M\o ller definitions and the calculated
expressions for energy. In Section 3 we perform a discussion of the results
and notice some particular cases. Through the paper we consider Latin indices
run from $0$ to $3$, geometrized units ($G=c=1$) and the signature ($+,-,-,-$).

\section{ENERGY\ OF\ REGULAR\ EXACT\ BLACK\ HOLE\ SOLUTIONS\ COUPLED\ WITH
NONLINEAR\ ELECTRODYNAMICS\ SOURCE}

\ In a recent study [14] E. Ay\'{o}n-Beato and A. Garcia developed a regular
class of four parametric exact black hole solutions of Einstein's equations
coupled with a nonlinear electrodynamics source. Notice that this class of
solutions describes regular exact solutions under some physically reasonable
assumptions. The nonlinear electrodynamics changes into the Maxwell theory in
the weak field approximation and the corresponding solutions behave
asymptotically as the Reissner-Nordstr\"{o}m solution. This class of solution
is described by four parameters, which are the mass $m$, the charge $q$, and a
sort of dipole and quadrupole moments of the nonlinear source $\alpha$ and
$\beta$, respectively. The $\alpha$ and $\beta$ parameters are determined by
the asymptotic behaviour of the electric field. Moreover, for some particular
range of the parameters the WEC energy condition is satisfied. Particular
imposed values of these parameters determine a previous solution given by the
authors [15].

The solution is given by%
\begin{equation}
ds^{2}=(1-\frac{2\,M(r)}{r})dt^{2}-(1-\frac{2\,M(r)}{r})^{-1}\,dr^{2}%
-r^{2}(d\theta^{2}+\sin^{2}\theta\,d\varphi^{2}), \tag{1}%
\end{equation}
where%
\begin{equation}
M(r)=\frac{m\,r^{\alpha}}{(r^{2}+q^{2})^{\alpha/2}}-\frac{q^{2}\,r^{\beta-1}%
}{2(r^{2}+q^{2})^{\beta/2}}. \tag{2}%
\end{equation}
For $\alpha\geq3$, $\beta\geq4$,\thinspace$\left\vert q\right\vert
\leq2\,s_{c}\,m$ (with $s=\left\vert q\right\vert /2\,m$ and $s_{c}$ the
critical value) these solutions describe regular charged black holes. The
geometry presents a similar global structure as the\ Reissner-Nordstr\"{o}m
solution, but the singularity at $r=0$ has been smoothed out and $r=0$
corresponds to the origin of spherical coordinates. The regular case
$\left\vert q\right\vert =2\,m$, $\alpha\geq1$, $\beta=\alpha+1$ does not
correspond to a black hole and satisfies the weak energy condition only for
the particular values $\alpha=1$ ($\beta=2$).

We compute the energy distribution of this class of solution with the
Einstein, Weinberg and M\o ller definitions. We present the Einstein, Weinberg
and M\o ller energy-momentum complexes [15] and the results obtained for the
energy distributions.  The energy component in the Einstein [1] prescription is 
given by \
\begin{equation}
E_{\mathrm{Einstein}}=\frac{1}{16\pi}\int\frac{\partial H_{0}^{\;\;0l}%
}{\partial x^{l}}d^{3}x, \tag{3}%
\end{equation}
where $H_{0}^{\;\;0l}$ is the corresponding von Freud superpotential
\begin{equation}
H_{0}^{\;\;0l}=\frac{g_{00}}{\sqrt{-g}}\frac{\partial}{\partial x^{m}}\left[
(-g)g^{00}g^{lm}\right]  . \tag{4}%
\end{equation}
For performing the calculations concerning the energy component of the
Einstein energy-momentum complex we have to transform the metric given by (1)
in the quasi-Cartesian coordinates $(t,x,y,z)$ and obtain
\begin{equation}
ds^{2}=Adt^{2}-(dx^{2}+dy^{2}+dz^{2})-\frac{A^{-1}-1}{r^{2}}(xdx+ydy+zdz)^{2}.
\tag{5}%
\end{equation}
In spherical coordinates the nonzero components of the Einstein
energy-momentum complex $H_{0}^{\;\;0l}$ are
\begin{equation}
H_{0}^{\;\;0r}=\frac{2\kappa}{r}\hat{r}-\frac{1}{A}\hat{r}(\hat{r}\cdot\nabla
A)+\frac{1}{A}\nabla A, \tag{6}%
\end{equation}
where we denote $\kappa=1-A$. Using the Gauss theorem we get
\begin{equation}
E_{\mathrm{Einstein}}=\frac{1}{16\pi}\oint H_{0}^{\;\;0r}\cdot\hat{r}%
r^{2}d\Omega, \tag{7}%
\end{equation}
and the integral being taken over a sphere of radius $r$, with the outward
normal $\hat{r}$ and the differential solid angle $d\Omega$. The expression
for energy within radius $r$ is given by
\begin{equation}
E_{\mathrm{Einstein}}=\frac{r}{2}(1-A)=m(1+\frac{q^{2}}{r^{2}})^{-\alpha
/2}-\frac{q^{2}}{2r}(1+\frac{q^{2}}{r^{2}})^{-\beta/2}. \tag{8}%
\end{equation}

The Weinberg energy-momentum complex [5] is defined as
\begin{equation}
\tau^{\nu\lambda}=\frac{\partial}{\partial x^{\rho}}Q^{\rho\nu\lambda},
\tag{9}%
\end{equation}
where $Q^{\rho\nu\lambda}$ is the Weinberg superpotential
\begin{equation}
Q^{\rho\nu\lambda}=\frac{\partial h_{\mu}^{\mu}}{\partial x_{\rho}}\eta
^{\nu\lambda}-\frac{\partial h_{\mu}^{\mu}}{\partial x_{\nu}}\eta^{\rho
\lambda}+\frac{\partial h^{\mu\nu}}{\partial x^{\mu}}\eta^{\rho\lambda}%
-\frac{\partial h^{\mu\rho}}{\partial x^{\mu}}\eta^{\nu\lambda}-\frac{\partial
h^{\nu\lambda}}{\partial x_{\rho}}+\frac{\partial h^{\rho\lambda}}{\partial
x_{\nu}}, \tag{10}%
\end{equation}
with $\eta_{\mu\nu}$ the Minkowski metric and $h_{\mu\nu}=g_{\mu\nu}-\eta
_{\mu\nu}$. The indices on $h_{\mu\nu}$ and $\partial/\partial x^{\lambda}$
can be raised and lowered with $\eta$. For the evaluation of the
energy-momentum in\ the Weinberg prescription we use the metric transformed in
quasi-Cartesian coordinates $(t,x,y,z)$ (5) together with the equation
\begin{equation}
P^{\lambda}=\frac{1}{16\pi}\int\frac{\partial Q^{i0\lambda}}{\partial x^{i}%
}d^{3}x. \tag{11}%
\end{equation}
The nonvanishing components $Q^{i00}$ obtained with the Weinberg
energy-momentum complex are given by
\begin{equation}
Q^{i00}=\frac{A^{\prime}}{r}\hat{r}+\frac{\hat{r}}{2}(\hat{r}\cdot\nabla
A^{\prime})-\frac{1}{2}\nabla A^{\prime}, \tag{12}%
\end{equation}
where $A^{\prime}=A^{-1}-1$. We apply the Gauss theorem and compute the energy
within radius $r$
\begin{equation}
E_{\mathrm{Weinberg}}=P^{0}=\frac{1}{16\pi}\oint Q^{i00}n_{i}r^{2}%
d\Omega=\frac{r}{2}A^{\prime}. \tag{13}%
\end{equation}
The connection between the energy component of the covariant energy-momentum
four vector of the Weinberg energy-momentum complex and the energy in the
Einstein\ prescription [15] is given by
\begin{equation}
E_{\mathrm{Weinberg}}^{covariant}=g_{00}E_{\mathrm{Weinberg}}=\frac{\kappa
r}{2}=E_{\mathrm{Einstein}} \tag{14}%
\end{equation}
and we have%
\begin{equation}
E_{\mathrm{Weinberg}}^{covariant}=E_{\mathrm{Einstein}}=\frac{r}%
{2}(1-A)=m(1+\frac{q^{2}}{r^{2}})^{-\alpha/2}-\frac{q^{2}}{2r}(1+\frac{q^{2}%
}{r^{2}})^{-\beta/2}. \tag{15}%
\end{equation}

Now, we perform the calculations in the M{\o }ller prescription. The M\o ller
energy-momentum complex~[7] is given by
\begin{equation}
\Theta_{\nu}^{\;\;\mu}=\frac{1}{8\pi}\frac{\partial\chi_{\nu}^{\;\;\mu\sigma}%
}{\partial x^{\sigma}},\tag{16}%
\end{equation}
where the antisymmetric M{\o}ller superpotential $\chi_{\nu}^{\;\;\mu\sigma}%
$\ is defined by
\begin{equation}
\chi_{\nu}^{\;\;\mu\sigma}=\sqrt{-g}\left(  \frac{\partial g_{\nu\alpha}%
}{\partial x^{\beta}}-\frac{\partial g_{\nu\beta}}{\partial x^{\alpha}%
}\right)  g^{\mu\beta}g^{\sigma\alpha}.\tag{17}%
\end{equation}
The expression for energy is computed as
\begin{equation}
E_{\mathrm{M{\o}ller}}=\frac{1}{8\pi}\int\frac{\partial\chi_{0}^{\;\;0k}%
}{\partial x^{k}}d^{3}x.\tag{18}%
\end{equation}
The M{\o }ller prescription allows to perform the calculations in any
coordinate system and the required nonzero component of the M{\o }ller
energy-momentum complex is
\begin{equation}
\chi_{0}^{\;\;0k}=r^{2}\sin\theta\frac{dA}{dr}.\tag{19}%
\end{equation}
The energy contained in a sphere of radius $r$ is given by
\begin{equation}
E_{\mathrm{M{\o }ller}}=\frac{r^{2}}{2}\frac{dA}{dr}=m(1+\frac{q^{2}}{r^{2}%
})^{-\alpha/2}(1-\frac{\alpha\,q^{2}}{r^{2}+q^{2}})-\frac{q^{2}}{2\,r}%
(1+\frac{q^{2}}{r^{2}})^{-\beta/2}(2-\frac{\beta\,q^{2}}{r^{2}+q^{2}%
}).\tag{20}%
\end{equation}
One notice the dependence of the energy in the three prescriptions on the mass
$m$, the charge $q$, and the parameters $\alpha$ and $\beta$, respectively.

In the foloowing we perform a study of the conditions that have to be
satisfied for obtaining the same expression for energy distribution using the
Einstein, Weinberg covariant and M\o ller energy-momentum complexes. We impose%
\begin{equation}
E_{\mathrm{Einstein}}=E_{\mathrm{Weinberg}}^{covariant}=E_{\mathrm{M{\o}ller}}
\tag{21}%
\end{equation}
and we obtain%
\begin{equation}
m(1+\frac{q^{2}}{r^{2}})^{-\alpha/2}(\frac{\alpha\,q^{2}}{r^{2}+q^{2}}%
)-\frac{q^{2}}{2\,r}(1+\frac{q^{2}}{r^{2}})^{-\beta/2}(-1+\frac{\beta\,q^{2}%
}{r^{2}+q^{2}})=0.\tag{22}%
\end{equation}
For the equation (22) we found the solutions $m=\frac{1}{2r\alpha}\left(
q^{2}+r^{2}\right)  \left(  q^{2}\frac{\beta}{q^{2}+r^{2}}-1\right)
\frac{\left(  \frac{q^{2}}{r^{2}}+1\right)  ^{\frac{1}{2}\alpha}}{\left(
\frac{q^{2}}{r^{2}}+1\right)  ^{\frac{1}{2}\beta}}$ and numeric $\beta
=24.\,\allowbreak432,r=-43.\,\allowbreak117,\alpha=-138.\,\allowbreak
89,m=1.\,\allowbreak054\,0\times10^{-24},q=165.\,\allowbreak14$. We conclude
that the energy calculated in Schwarzschild cartesian coordinates with the
Einstein and Weinberg covariant energy-momentum complexes is equal to the
energy obtained in the M\o ller prescription in this particular cases.

\section{DISCUSSION}

Despite the weakness of the energy-momentum complexes, which is connected to
their coordinate dependence excepting the M\o ller prescription, many
interesting studies have been performed in the recent years. Physicists like
Bondi [16], Misner [17] and Lessner [18] showed that the pseudotensorial
definitions could be a reliable solution (starting point) for solving the
energy-momentum localization issue.

Our work is focused on the evaluation of the energy distribution of a regular
class of four parametric exact black hole solutions of Einstein's equations
coupled with a nonlinear electrodynamics source given by Ay\'{o}n-Beato and
Garc\'{\i}a. We found that the expressions for energy yielded by the Einstein,
Weinberg and M\o ller prescriptions depend on the mass $m$, the charge $q$,
and $\alpha$ and $\beta$ that represent a sort of dipole and quadrupole
moments of the nonlinear source, respectively. Further, the Weinberg covariant
energy distribution is the same as the Einstein energy. Also, we extend a
previous work [15] and our results contain this as a particular case. In
addition, we make a discussion concerning the conditions which have to be
satisfied for obtaining the same expression for energy using the Einstein,
Weinberg covariant and M\o ller energy-momentum complexes. 

We present the expressions for energy obtained in these prescriptions in the
following table. In addition some particular cases are discussed, and here
belong the cases of the Reissner-Nordstr\"{o}m solution (vanishing values for
$\alpha$ and $\beta$) and Schwarzschild solution.

Our conclusions are:

a) The expressions for energy are finite in all definitions, excepting the
particular case $\alpha=\beta=0,$ $r\rightarrow0$, which gives the same
infinite result for energy distributions in the Einstein, Weinberg and
M\o ller definitions.

b) For $\alpha=3$, $\beta=4$ we find the case of the regular black hole
solution given by Ay\'{o}n-Beato and Garc\'{\i}a [19] and studied by the
authors [15].

c) We observe that the particular cases $q\rightarrow0$, $\alpha=\beta
=0$,$\,q\rightarrow0$ and $\alpha=\beta=0,\,r\rightarrow\infty$ yield the same
result for energy, in all the prescriptions the energy of the black hole is
equal to the ADM mass, and this result also corresponds to the Schwarzschild solution.

d) The special case $\alpha\geq3$, $\beta\geq4$, $\left\vert q\right\vert
\leq2\,s_{c}\,m$, $r\rightarrow0$ gives a zero value for energy in all the
prescriptions, and even the expressions for energy are the same this result
doesn't lead to a physically meaningful interpretation.

e) Excepting the first two cases from the table below, there is a total
compatibility between the results obtained with the Einstein, Weinberg and
M\o ller definitions.%

\[%
\begin{tabular}
[c]{|l|l|l|}\hline
Case & $E_{\mathrm{Einstein}}=E_{\mathrm{Weinberg}}^{covariant}$ &
$E_{\mathrm{M{\o}ller}}$\\\hline%
\begin{tabular}
[c]{l}%
$\alpha\geq3$, $\beta\geq4$,\thinspace\\
$\left\vert q\right\vert \leq2\,s_{c}\,m$%
\end{tabular}
&
\begin{tabular}
[c]{l}%
$m(1+\frac{q^{2}}{r^{2}})^{-\alpha/2}-$\\
$-\frac{q^{2}}{2r}(1+\frac{q^{2}}{r^{2}})^{-\beta/2}$%
\end{tabular}
& $%
\begin{tabular}
[c]{l}%
$m(1+\frac{q^{2}}{r^{2}})^{-\alpha/2}(1-\frac{\alpha\,q^{2}}{r^{2}+q^{2}})-$\\
$-\frac{q^{2}}{2\,r}(1+\frac{q^{2}}{r^{2}})^{-\beta/2}(2-\frac{\beta\,q^{2}%
}{r^{2}+q^{2}})$%
\end{tabular}
\ \ \ $\\\hline
$\alpha=3$, $\beta=4$ &
\begin{tabular}
[c]{l}%
$m(1+\frac{q^{2}}{r^{2}})^{-3/2}-$\\
$-\frac{q^{2}}{2r}(1+\frac{q^{2}}{r^{2}})^{-2}$%
\end{tabular}
& $%
\begin{tabular}
[c]{l}%
$m(1+\frac{q^{2}}{r^{2}})^{-5/2}(1-\frac{2q^{2}}{r^{2}})-$\\
$-\frac{q^{2}}{r}(1+\frac{q^{2}}{r^{2}})^{-3}(1-\frac{q^{2}}{r^{2}})$%
\end{tabular}
\ \ \ $\\\hline
$q\rightarrow0$ & $m$ & $m$\\\hline
$\alpha=\beta=0$,$\,$($RN$) & $m-\frac{q^{2}}{2r}$ & $m-\frac{q^{2}}{r}%
$\\\hline%
\begin{tabular}
[c]{l}%
$\alpha\geq3$, $\beta\geq4$,\\
$\left\vert q\right\vert \leq2\,s_{c}\,m$,\\
$r\rightarrow0$%
\end{tabular}
& $0$ & $0$\\\hline
$\alpha=\beta=0$,$\,q\rightarrow0$ & $m$ & $m$\\\hline
$\alpha=\beta=0,r\rightarrow0$ & $\pm\infty$ & $\pm\infty$\\\hline
$\alpha=\beta=0,\,r\rightarrow\infty$ & $m$ & $m$\\\hline
\end{tabular}
\ \ \ \
\]

The equality of the results in the Einstein and Weinberg prescriptions, and in
addition the equality of the results in all prescriptions in some particular
cases demonstrate that the energy-momentum complexes can yield consistent
expressions for energy and are useful methods for energy-momentum localization.

\end{document}